\documentstyle[12pt]{article}

\begin{document}
\begin{center}
\large{{\bf Frobenius theorem and invariants for Hamiltonian systems}}
\vskip 5mm

F. Haas \\ Laboratoire de Physique des Milieux Ionis\'es, Universit\'e Henri Poincar\'e \\
BP 239, 54506 Vandoeuvre-les-Nancy, France
\end{center}
\vskip 5mm

\noindent
{\bf Abstract}

\noindent
We apply Frobenius integrability theorem in the search of invariants
for o\-ne-di\-men\-sio\-nal Hamiltonian systems with a time-dependent
potential. We obtain several classes of potential functions for which
Frobenius theorem assures the existence of a two-dimensional foliation to
which the motion is constrained. In particular, we derive a new
infinite class of potentials for which the motion is assurately restricted
to a two-dimensional foliation. In some cases, Frobenius theorem allows
the explicit construction of an associated invariant. It is proven the
inverse result that, if an invariant is known, then it
always can be furnished by Frobenius theorem.


\newpage

\section{Introduction}

The search for invariants (also called constants of motion or first
integrals) for dynamical systems is a classical topic in nonlinear science,
and in the past two decades there has been an intensive research on the
subject. A comprehensive review of the recent advances on the field can be
find in \cite{Ref1}. There are several techniques for the construction of
invariants, among which the direct method \cite{Ref2}, application of Noether's
theorem \cite{Ref3}, utilization of Lie symmetries \cite{Ref4},
Painlev\'e's analysis \cite{Ref5} and Darboux method \cite{Ref6}.
None of these methods has an universal character, and in most cases one or
more {\it ad hoc} assumption has to be made for obtaining concrete results.
For instance, in the case of Noether and Lie's approach an useful
simplification is achieved by considering point symmetries only \cite{Ref7,
Ref77}.

In the present work, we focus our attention on Hamiltonian systems with a
time-dependent Hamiltonian function of the form
\begin{equation}
\label{eq1}
H(q,p,t) = \frac{1}{2}p^2 + V(q,t) \,,
\end{equation}
where $V(q,t)$ is a time-dependent potential function and $p$ and $q$ are
coordinates on a two-dimensional phase space. Hamiltonians of type
(\ref{eq1}) appear in several branches of physics such as plasma physics
and quantum mechanics. Here, we are interested in particular in the
construction of invariants for suitable classes of potentials. By invariant,
we understand precisely a function $I = I(q,p,t)$ globally defined and having the property
\begin{equation}
\label{eq2}
\dot I = \frac{\partial I}{\partial t} + p\frac{\partial I}{\partial q} - 
\frac{\partial V}{\partial q}\frac{\partial I}{\partial p} = 0 \,,
\end{equation}
where the dot represents time differentiation. In other words, equation
(\ref{eq2}) says that the function $I$ is constant along the trajectories of
the canonical equations of motion. The existence of an invariant immediately
prevents the appearance of chaos, since when a first integral is available
the motion is restricted to a two-dimensional surface $I(q,p,t) =$ cte. As is 
well known, if the independent variable is continuous, chaos may take place
only in dynamical systems with three or more dimensions. Note that an
explicit time-dependence of $I$ does not modify at all the fact that chaos
is impossible if an invariant is known to exist (provided $I(q,p,t) =$ cte.
defines a sufficiently smoth surface).

To begin our analysis, we write the dynamical vector field associated to 
the canonical equations of motion,
\begin{equation}
\label{eq3}
{\bf u} = \frac{\partial}{\partial t} + p\frac{\partial}{\partial q} - 
\frac{\partial V}{\partial q}\frac{\partial}{\partial p} \,.
\end{equation}
Now, the definition (\ref{eq2}) reads ${\bf u}(I) = 0$. The dynamical
vector field ${\bf u}$ is defined in a three-dimensional space, with
coordinates $q$, $p$ and $t$. Incidentally, some years ago
\cite{Ref8, Ref9} it was introduced a method for the derivation of
invariants for three-dimensional dynamical systems. This method relies on
Frobenius integrability theorem, and has been used in the analysis of the
three-dimensional Lotka-Volterra system and the May-Leonard and Lorenz
systems \cite{Ref8, Ref9}. Presently, we apply Frobenius method in order to
derive integrable classes of Hamiltonians of the form (\ref{eq1}). The large
number of integrable three-dimensional Lotka-Volterra systems detected by
Frobenius method is a good reason for optimism. Indeed, as seen in the
following, we are able to find several integrable potentials by using
Frobenius method with some auxiliary hypothesis. In particular, we find a
new whole infinite family of potentials for which the motion is shown to be
restricted to a two-dimensional surface. Finally, we prove the result that,
if an invariant is known to exist, then it can be derived by Frobenius
method. This last result shows that the Frobenius approach for invariants
for one-dimensional time-dependent Hamiltonian systems necessarily
reproduces all known integrable cases in the literature. 

The article is organized as follows. In Section II, we describe Frobenius
method and derive the basic equation to be solved in the case of Hamiltonian
systems with Hamiltonian of the form (\ref{eq1}). In Section III, we use a
linear {\it ansatz} in momentum for obtaining particular solutions of the
basic equation derived in Section II. As a result, we derive the invariant
linear in momentum for the forced time-dependent harmonic oscillator
\cite{Ref10} and Sarlet's invariant \cite{Ref11}-\cite{Ref13}.
In Section IV, we consider functions rational in momentum, so
obtaining an infinite family of potentials amenable to Frobenius method.
As particular cases in this family, we found the potentials having an
invariant quadratic in momentum \cite{Ref10} and Giacomini potentials
\cite{Ref14, Ref15}. Also we show a class of ``weakly integrable''
potentials, which have not in general an invariant but for which Frobenius
method works. Section V is reserved to the conclusions.

\section{Frobenius method and basic equation}

The Frobenius method was already exposed in \cite{Ref8}.
Here, we only show the essentials of the procedure, with a view to
application to one-dimensional time-dependent
Hamiltonian systems. The basic aim in the Frobenius procedure for the
construction of invariants of motion is to find a vector field ${\bf v}$
given by
\begin{equation}
\label{eq4}
{\bf v} = A\frac{\partial}{\partial t} + B\frac{\partial}{\partial q} +  
C\frac{\partial}{\partial p} 
\end{equation}
which is compatible with the dynamical vector field. In equation
(\ref{eq4}), $A = A(q,p,t)$, $B = B(q,p,t)$ and $C = C(q,p,t)$ are
functions to be determined. Compatibility
means that 
\begin{equation}
\label{eq5}
[{\bf u},{\bf v}] = \alpha{\bf u} + \beta{\bf v} \,,
\end{equation}
where $[,]$ represents the Lie bracket, $[{\bf u},{\bf v}] = {\bf u}{\bf v}
- {\bf v}{\bf u}$, and $\alpha = \alpha(q,p,t)$ and $\beta = \beta(q,p,t)$
are functions (not
necessarily constants) on phase space and time. Equation (\ref{eq5})
implies that the
dynamical vector field, the compatible vector field and their Lie bracket
are linearly
dependent. If $\beta = 0$, the compatibility condition (\ref{eq5}) reduces
to the statement that ${\bf v}$ is a generator of Lie symmetries for the
dynamical system \cite{Ref4}. For $\beta \neq 0$, there is a generalization.

If a compatible vector field can be found, then the Frobenius theorem
assures the
existence of a maximal connected leaf which is tangent to both ${\bf u}$
and ${\bf v}$, in
each point in the three-dimensional space with coordinates $(q,p,t)$. 
Moreover, the union of all such leaves provides a foliation, that is, each
point $(q,p,t)$
is contained in one and only one leaf. However, this two-dimensional
foliation is not necessarily defined in terms of the level surfaces of some
function $I(q,p,t)$. Nevertheless, for the derivation of an invariant common
to the dynamical and the compatible vector fields we assume the existence of
such function, so that $I(q,p,t) =$ cte. provides a representation of the
foliation.  The fact that ${\bf u}$ and ${\bf v}$ are tangent to the maximal
connected leaf implies that ${\bf u}(I) = {\bf v}(I) = 0$.

Suppose that a compatible vector field ${\bf v}$ is available. The natural
question in this
circumstance is how to construct an associated invariant $I(q,p,t)$. The
procedure for the derivation of an invariant in Frobenius method is
identical to the construction of first integrals by Lie symmetry methods
\cite{Ref4}. By definition, ${\bf u}(I) = {\bf v}(I) = 0$, since the
maximal surface is tangent to both the dynamical and the compatible vector
field. Consider
\begin{equation}
\label{eq6}
{\bf v}(I) = 0 \,.
\end{equation}
Denoting the integral surfaces of the compatible vector field by 
$f(q,p,t) =$ cte., $g(q,p,t) =$ cte., so that ${\bf v}(f) = {\bf v}(g) = 0$,
we have that
\begin{equation}
\label{eq7}
I = I(f,g) 
\end{equation}
is the general solution to (\ref{eq6}). As shown in \cite{Ref8},
substitution of the form (\ref{eq7}) in the condition ${\bf u}(I) = 0$
leads to a first-order ordinary differential equation,
\begin{equation}
\label{eq8}
\frac{df}{dg} = \Lambda(f,g) \,,
\end{equation}
where $\Lambda$ is a function depending on $f$ and $g$ only. 
Expressing the constant of integration of the later equation in terms of
$f(q,p,t)$ and $g(q,p,t)$, we obtain the invariant associated to the
compatible vector field. As seen by the reasoning in this paragraph, the
Lie and Frobenius strategies for derivation of invariants are indeed equal,
once ${\bf v}$ is known. The sole difference between the two methods is that
the compatible vector field is not necessarily a generator of Lie symmetries.

Equation (\ref{eq6}) is a first-order linear partial differential equation
which has to be
solved by the method of characteristics. One may expects, a priori, that
this equation is more
easy to handle than the original dynamical system. Otherwise, existence of a
compatible vector field only assures
that the motion is restricted to a two-dimensional foliation, but does not
provides the explicit form of an associated invariant. Another possible
drawback in the procedure is the resolution of (\ref{eq8}). Systems
having a compatible vector field but for which one is not able to find
explicitly the associated constant of motion were termed weakly integrable
in the literature \cite{Ref8}. In Section IV, we show an infinity family of
potential functions which pertains to such weakly integrable category.

Turning our attention to our specific problem, the study of one-dimensional 
time-dependent Hamiltonian systems, let us consider the dynamical and
compatible vector fields  (\ref{eq3}-\ref{eq4}). Actually, there is no
loss of generality in taking $A = 0$ and $B = 1$ in the definition of the
compatible vector field. Indeed, suppose that the compatibility condition
(\ref{eq5}) is fulfilled. Defining a new vector field ${\bf v}'$ by
\begin{equation}
\label{eq9}
{\bf v}' = \mu{\bf u} + \nu{\bf v} \,,
\end{equation}
we get, from (\ref{eq5}), 
\begin{equation}
\label{eq11}
[{\bf u},{\bf v}'] = \left({\bf u}(\mu) + \alpha\nu - \mu\left(\frac{{\bf u}(\nu)}{\nu} + \beta\right)\right){\bf u} + \left(\frac{{\bf u}(\nu)}{\nu} + \beta\right){\bf v}' \,.
\end{equation}
Thus, ${\bf u}$ and ${\bf v}'$ are compatible. Now, define 
\begin{equation}
\label{eq10}
\mu = \frac{A}{Ap - B} \,, \quad \nu = \frac{1}{B - Ap} \,.
\end{equation}
We always take $B \neq Ap$, so that there is no singularity in the
denominator of $\mu$ and $\nu$ as given in (\ref{eq10}). The case $B = Ap$
can be shown to give only trivial results (dynamical and compatible
vector fields linearly dependent).  With the choice (\ref{eq10}), we have
\begin{equation}
\label{eq12}
{\bf v}' = \frac{\partial}{\partial q} + \left(\frac{C + A\partial\,V/\partial\,q}{B - Ap}\right)\frac{\partial}{\partial p} \,.
\end{equation}
For all compatible vector fields ${\bf v}$, we can construct an associated
compatible vector field ${\bf v}'$ as defined by (\ref{eq12}). In
conclusion, vector fields of the form
\begin{equation}
\label{eq13}
{\bf v} = \frac{\partial}{\partial q} + C(q,p,t)\frac{\partial}{\partial p}
\end{equation}
are all that is needed in our calculations. There is no real gain in
considering compatible vector fields more general than (\ref{eq13}).
Observe that the use of compatible vector fields like (\ref{eq13}) is
analogous to the use of generators in evolutionary form, in the theory of
Lie extended groups \cite{Ref4}. However, here there is even more
simplification, since the coefficient of $\partial/\partial\,q$ in ${\bf v}$
was set to unity. Finally, a major advantage of using the reduced form
(\ref{eq13}) is that the integral surface $g = t =$ cte. of the compatible
vector field is known in advance. There remains only the task of computing
the second integral surface.

Let us study the consequences of the compatibility condition. Computing the
Lie bracket of the dynamical vector field and ${\bf v}$ as given in
(\ref{eq13}), we find
\begin{equation}
\label{eq14}
[{\bf u},{\bf v}] = - C\frac{\partial}{\partial q} +
\left({\bf u}(C) + \frac{\partial^{2}V}{\partial\,q^2}\right)\frac{\partial}{\partial p} \,.
\end{equation} 
Now using the compatibility condition (\ref{eq5}), we determine the
coefficients $\alpha$ and
$\beta$, 
\begin{equation}
\label{eq15}
\alpha = 0 \,, \quad \beta = - C \,. 
\end{equation}
In general $\beta$ is not zero, so that the compatible vector field is not a
generator of Lie symmetries.

The compatibility requirement (\ref{eq5}) comprises three equations, since
we are dealing with vector fields in three dimensions. Two of these
equations were used to calculate $\alpha$ and $\beta$. The third equation
gives
\begin{equation}
\label{eq16}
{\bf u}(C) + C^2 = - \frac{\partial^{2}V}{\partial\,q^2} \,.
\end{equation}
The latter equation may be used to determine simultaneously the potential
and the function $C$, and is the basic tool in Frobenius method for
one-dimensional
Hamiltonian systems with a time-dependent potential. In the next sections,
we show several  solutions for (\ref{eq16}). To deal with (\ref{eq16}) we
have to suppose appropriate particular forms for the function $C$. As
remarked in the introduction, most methods of derivation of invariants are
not free from some assumption. Frobenius approach is not an exception.

To end this section, we comment on the possible equivalence between the
methods of Lie and
Frobenius for the construction of invariants. Indeed, if a compatible
vector field ${\bf v}$ is known, one may wonder if a generator
${\bf v}'$ of Lie symmetries may be obtained. So, consider ${\bf v}'$ of
the form (\ref{eq9}). Frobenius method generalizes Lie's approach by
allowing a non zero $\beta$ function in equation (\ref{eq5}). A look at
equation (\ref{eq11}) shows that we have a Lie symmetry if we choose $\nu$
so that
\begin{equation}
\label{eq17}
{\bf u}(\nu) + \beta\nu = 0 \,.
\end{equation}
Any vector field ${\bf v}'$ of the form (\ref{eq9}), with arbitrary $\mu$
and with $\nu$
satisfying (\ref{eq17}), is a generator of Lie symmetries provided ${\bf v}$
is a compatible vector field. However, the difficult point
in this reasoning is that (\ref{eq17}) is a first-order linear partial
differential for
$\nu$, which can be solved in general only if the integral surfaces of
the dynamical
vector field are known. Obviously, {\it a priori} this information is not
available, at least for nontrivial systems.

In Sections III and IV, we solve the fundamental equation (\ref{eq16}) for
selected functional forms of $C(q,p,t)$. We use $C$ in the form of functions
linear and rational
in momentum. These two cases are treated separately.

\section{$C(q,p,t)$ linear in momentum}

As remarked earlier, to obtain definite results some assumption must be
made on the functional dependence of $C(q,p,t)$. In this section, we
consider
\begin{equation}
\label{eq18}
C = C_{0} + C_{1}p \,,
\end{equation}
where $C_{0} = C_{0}(q,t)$ and $C_{1} = C_{1}(q,t)$ are functions depending
only on position and time. Inserting the linear {\it ansatz} (\ref{eq18})
into (\ref{eq16}), there results that a quadratic polynomial in momentum
must be identically zero. Equating to zero the coefficients of the various
powers of $p$, we find a system of partial differential equations,
\begin{eqnarray}
\label{eq19}
\frac{\partial C_1}{\partial q} + C_{1}^2  &=& 0 \,,\\
\label{eq20}
\frac{\partial C_1}{\partial t} + \frac{\partial C_0}{\partial q} + 2\,C_{0}C_{1} &=& 0 \,, \\
\label{eq21}
\frac{\partial C_0}{\partial t} + C_{0}^2 - C_{1}\frac{\partial V}{\partial q} + \frac{\partial^{2}V}{\partial q^2} &=& 0 \,. 
\end{eqnarray}
This system is nonlinear. To solve it, first observe that (\ref{eq19})
admits two types of solution,
\begin{equation}
\label{eq22}
C_1 = 0 \,,
\end{equation}
and 
\begin{equation}
\label{eq23}
C_1 = \frac{1}{q - \sigma} \,,
\end{equation}
where $\sigma = \sigma(t)$ is an arbitrary function. The two branches
(\ref{eq22}-\ref{eq23}) will be studied separately.

\subsection{Linear invariants}

Let us first consider the case $C_1 = 0$. In this circumstance,
(\ref{eq20}) may be readily solved, yielding
\begin{equation}
\label{eq24}
C_0 = \frac{\dot\rho}{\rho} \,,
\end{equation}
where $\rho = \rho(t) \neq 0$ is an arbitrary function. The form
(\ref{eq24}) was chosen for later convenience. We can now solve for
the potential using (\ref{eq21}), obtaining
\begin{equation}
\label{eq25}
V = V_{0}(t) - \frac{\dot{F}(t)q}{\rho} - \frac{\ddot\rho\,q^2}{2\rho} \,,
\end{equation}
where $V_{0}(t)$ and $F(t)$ are new arbitrary functions of time. In fact,
we can set $V_0 = 0$ without loss of generality, since it does not
contribute to the equations of motion.

The potential (\ref{eq25}) corresponds to the forced time-dependent
harmonic oscillator system. To construct an invariant for it using
Frobenius method, consider the associated compatible vector field, which,
according to (\ref{eq13}) and (\ref{eq18}), reads
\begin{equation}
\label{eq26}
{\bf v} = \frac{\partial}{\partial q} + \frac{\dot\rho}{\rho}\frac{\partial}{\partial p} \,.
\end{equation}
The integral surfaces of ${\bf v}$ are specified by 
\begin{equation}
\label{eq27}
f = p - \frac{\dot\rho\,q}{\rho} \,, \quad g = t \,.
\end{equation}
Computing $df/dt$ along the trajectories of the dynamical system, we obtain
\begin{equation}
\label{eq28}
\frac{df}{dg} = - \frac{\dot\rho\,f}{\rho} + \frac{\dot F}{\rho} \,.
\end{equation}
As $\rho = \rho(g)$, the right hand side of the last equation is indeed a
function of $f$ and $g$ only, in accordance with the general result
(\ref{eq8}). Solving (\ref{eq28}), we obtain the well known \cite{Ref10}
invariant linear in momentum for the forced time-dependent harmonic
oscillator,
\begin{equation}
\label{eq29}
I = \rho\,p - \dot\rho\,q - F \,.
\end{equation}

\subsection{Sarlet's potential}

The previous calculations show how the Frobenius method works. Let us
consider the less trivial solution (\ref{eq23}) for (\ref{eq19}) and
proceed to the solution of (\ref{eq20}-\ref{eq21}).
Taking into account equations (\ref{eq23}) and (\ref{eq20}), we
obtain
\begin{equation}
\label{eq30}
C_0 = - \frac{\dot\sigma(q - \sigma) + 2\gamma}{(q - \sigma)^2} \,,
\end{equation}
where $\gamma = \gamma(t)$ is an arbitrary function of time. With $C_0$
and $C_1$ specified by (\ref{eq23}) and (\ref{eq30}), we can find the
potential using (\ref{eq21}). It reads
\begin{equation}
\label{eq31}
V = V_{0}(t) - \ddot\sigma(q - \sigma) - \frac{\ddot\rho\,q^2}{2\rho} - \frac{\gamma^2}{2(q - \sigma)^2} - \dot\gamma\log(q - \sigma) \,, 
\end{equation}
introducing the new arbitrary functions $V_0 = V_{0}(t)$
and $\rho = \rho(t) \neq 0$.

Having solved the system (\ref{eq19}-\ref{eq21}), there remains the task of
obtaining the associated invariant using the compatible vector field, which
is in the present case
\begin{equation}
\label{eq32}
{\bf v} = \frac{\partial}{\partial q} + \frac{(q - \sigma)(p - \dot\sigma) - 2\gamma}{(q - \sigma)^2}\,\frac{\partial}{\partial p} \,.
\end{equation}
The associated invariant surfaces are specified by
\begin{equation}
\label{eq33}
f = \frac{p - \dot\sigma}{q - \sigma} - \frac{\gamma}{(q - \sigma)^2} \,, \quad g = t \,.
\end{equation}
Along the canonical equations of motion with potential (\ref{eq31}), we
obtain
\begin{equation}
\label{eq34}
\frac{df}{dg} = - f^2 + \frac{\ddot\rho}{\rho} \,,
\end{equation}
a Riccati equation. Note that the general result (\ref{eq8}) is indeed
verified, since $\rho = \rho(g)$. The general solution for (\ref{eq34}) is
\begin{equation}
\label{eq35}
f = \frac{\dot\rho}{\rho} - \frac{1}{\rho^{2}\left(I - \int^{t}dt'/\rho^{2}\right)} \,,
\end{equation}
where $I$ is the constant of integration for the Riccati equation.
Actually, $I$ is the invariant of the problem, and, using (\ref{eq33})
and (\ref{eq35}), we have
\begin{equation}
\label{eq36}
I = \int^{t}\frac{dt'}{\rho^2} - \frac{(q - \sigma)/\rho}{\rho(p - \dot\sigma) - \dot\rho(q - \sigma) - \gamma\rho/(q - \sigma)} \,.
\end{equation}
The potential (\ref{eq31}) and the invariant (\ref{eq36}) are not new,
being derived by the first time by Sarlet and then by a variety of methods
\cite{Ref11}-\cite{Ref13}.

In this section, we have shown how Frobenius procedure works, in the
elementary case of functions $C(q,p,t)$ that are linear in momentum. In
this way, we have obtained some already known results. In order to find
new integrable or weakly integrable potentials, a more complicated momentum
dependence of $C(q,p,t)$ must be utilized. It can be easily shown that
higher order polynomial forms of $C(q,p,t)$ does not gives noting new.
Hence, we proceed differently in the next section, taking a rational form
for $C(q,p,t)$.

\section{$C(q,p,t)$ rational in momentum}

In this section we take the following {\it ansatz} for the solution of the
basic equation (\ref{eq16}),
\begin{equation}
\label{eq60}
C = \frac{C_0 + C_{1}p}{p - C_2} \,,
\end{equation}
where $C_0$, $C_1$ and $C_2$ are functions of coordinate and time only. We
also assume that $C_0 + C_{1}C_{2} \neq 0$, so
that $\partial\,C/\partial\,p \neq 0$. Substitution of (\ref{eq60})
into (\ref{eq16}) yields an equation implying that a cubic polynomial
in momentum is identically zero. Again, the coefficient of equal powers in
momentum must be zero, resulting in a system of partial differential
equations for $C_0$, $C_1$, $C_2$ and the potential. Considering the
coefficient of $p^3$, we get that $C_1$ is a function of time only. For
latter convenience, we represents the solution as
\begin{equation}
\label{eq61}
C_1 = \dot\rho/\rho \,,
\end{equation}
where $\rho = \rho(t)$ is an arbitrary nonzero function of $t$. Taking into
account (\ref{eq61}), one can show that the term proportional to $p^2$ in
the basic system of equations yields
\begin{equation}
\label{eq62}
C_0 = - \ddot\sigma - \frac{\ddot\rho}{\rho}(q - \sigma) - \frac{\dot\rho}{\rho}\,C_2 
- \frac{\partial V}{\partial q} \,,
\end{equation}
where $\sigma = \sigma(t)$ is a new arbitrary function of time. Equation
(\ref{eq62}) express  $C_0$ in terms of $C_2$ and $V$.

The last equations, corresponding to the linear term in momentum and the
remaining term, involves $C_1$ and $V$. After some algebra, we can
transform this system into an equivalent one, consisting of an equation for
$C_2$ only,
\begin{equation}
\label{eq63}
\frac{\partial C_2}{\partial t} + C_{2}\frac{\partial C_2}{\partial q} = \ddot\sigma + \frac{\ddot\rho}{\rho}(q - \sigma)  \,,
\end{equation}
and an equation determining the potential, 
\begin{eqnarray}
\frac{\partial V}{\partial t} + C_{2}\frac{\partial V}{\partial q} &=&
- \frac{2\dot\rho}{\rho}\,V + (\sigma{\buildrel\cdots\over\rho} +
(\sigma\dot\rho + \rho\dot\sigma)\ddot\rho/\rho -
2\dot\rho\ddot\sigma -
\rho{\buildrel\cdots\over\sigma})\frac{q}{\rho} \nonumber \\
\label{eq64}
&-& (\frac{{\buildrel\cdots\over\rho}}{\rho}
+ \frac{\dot\rho\ddot\rho}{\rho^2})\frac{q^2}{2}
- (\frac{\ddot\rho}{\rho}(q - \sigma) + \ddot\sigma)\,C_2 \\
&+& \frac{1}{\rho^2}\frac{d}{dt}(\rho^{2}V_{0}(t)) \,, \nonumber
\end{eqnarray}
where $V_{0}(t)$ is an arbitrary function of time. 

The strategy for solving (\ref{eq63}-\ref{eq64}) is clear. First, we have
to solve (\ref{eq63}), obtaining $C_2$ in terms of $(q,t)$. Then,
inserting this solution into (\ref{eq64}), we arrive at a well defined
partial differential equation for the potential. If its solution is
available, we can obtain $C_0$ and the corresponding associated compatible
vector field using (\ref{eq62}).

Fortunately, the solution for equation (\ref{eq63}) is available \cite{Ref16}
for arbitrary $\rho$ and $\sigma$, and reads
\begin{equation}
\label{eq65}
C_2 = \frac{1}{\rho}\left(\dot\rho(q - \sigma) +
\rho\dot\sigma + Q\right) \,,
\end{equation}
where $Q = Q(q,t)$ is implicitly defined according to
\begin{equation}
\label{eq66}
Q = F\left(\frac{q - \sigma}{\rho} - Q\int^{t}\frac{dt'}{\rho^2}\right) \,,
\end{equation}
where $F$ is an arbitrary function of the indicated argument. As $F$
depends on $Q$, the solution has indeed an implicit character. However, the
implicit function theorem assures that  we can always locally solve for
$Q(q,t)$ under suitable conditions on $F$. With $Q(q,t)$, we can write $C_2$
using (\ref{eq65}) and then the equation (\ref{eq64}) for the potential.
Note that the appearance of implicit relations is not new in the theory of
integrable systems (See, e.g., \cite{Ref14}, \cite{Ref16}).  

Taking into account the results of the section, we can write the compatible
vector field as
\begin{equation}
\label{eq67}
{\bf v} = \frac{\partial}{\partial q} + \left(\frac{\dot\rho}{\rho} - \frac{\rho(\partial\,V/\partial\,q + \ddot\sigma) + \ddot\rho(q - \sigma)}{\rho(p - \dot\sigma) - \dot\rho(q - \sigma) - Q}\right)\frac{\partial}{\partial p} \,.
\end{equation}
To obtain explicitly the compatible vector field, we have to define the
function $F$ in (\ref{eq66}) and the find the potential solving
(\ref{eq64}). Remembering that the latter is a linear partial differential
equation with two independent variables, we conclude that an additional
arbitrary function will appear after solving (\ref{eq64}) by
characteristics. In the continuation, we illustrate the whole procedure
with specific examples.

\subsection{Quadratic invariants}

Let us first consider the case of invariants quadratic in momentum. For this
class of solutions,
we take $F = 0$ in (\ref{eq66}). According to (\ref{eq65}), we then have
\begin{equation}
\label{eq68}
C_2 = \dot\sigma + \frac{\dot\rho}{\rho}(q - \sigma)  \,.
\end{equation}
Inserting this form into (\ref{eq64}), we obtain the equation determining
the potential. The solution is
\begin{equation}
\label{eq69}
V = V_{0}(t) + (\sigma\ddot\rho - \rho\ddot\sigma)\frac{q}{\rho}
- \frac{\ddot\rho\,q^2}{2\rho}
+ \frac{1}{\rho^2}U\left(\frac{q - \sigma}{\rho}\right) \,,
\end{equation}
where $U$ is an arbitrary function of the indicated argument. This is an
example of how the solution of the equation (\ref{eq64}) by characteristics
yield an additional arbitrary function in the complete solution. According
to (\ref{eq67}), the compatible vector field is
\begin{equation}
\label{eq70}
{\bf v} = \frac{\partial}{\partial q} +
\left(\frac{\dot\rho}{\rho} -
\frac{dU/d\bar{q}}{\rho^{2}(\rho(p - \dot\sigma) - \dot\rho(q
- \sigma))}\right)\frac{\partial}{\partial p} \,,
\end{equation}
where 
\begin{equation}
\label{eq71}
\bar{q} = \frac{q - \sigma}{\rho} \,.
\end{equation}

Equation (\ref{eq69}) indeed defines the class of potentials admitting an
invariant quadratic in momentum \cite{Ref10}.
To proceed with Frobenius method for the derivation of invariants, we use
a new momentum variable
\begin{equation}
\label{eq72}
\bar{p} = \rho(p - \dot\sigma) - \dot\rho(q - \sigma) \,,
\end{equation}
so that the compatible vector field reads
\begin{equation}
\label{eq73}
{\bf v} = \frac{1}{\rho\bar{p}}\left(\bar{p}\frac{\partial}{\partial\bar{q}}
- \frac{dU}{d\bar{q}}\frac{\partial}{\partial\bar{p}}\right) \,.
\end{equation}
The integral surfaces of ${\bf v}$ are 
\begin{equation}
\label{eq200}
f = \frac{\bar{p}^2}{2} + U(\bar{q}) \,, \quad g = t \,.
\end{equation}
This yields directly the energy-like invariant
\begin{equation}
\label{eq74}
I = \frac{\bar{p}^2}{2} + U(\bar{q}) \,,
\end{equation}
since, incidentally, $df/dg = 0$ along the trajectories. But $I$ 
is the invariant quadratic in momentum derived in \cite{Ref10},
expressed in terms of transformed coordinate and momentum. Actually,
(\ref{eq71}-\ref{eq72}) is an example of generalized canonical
transformation \cite{Ref10}. In the present context, $I$ as given in
(\ref{eq74}) corresponds to an integral surface of the compatible vector
field.

\subsection{Giacomini potentials}

As another particular choice of arbitrary functions, consider 
\begin{equation}
\label{eq90}
\rho = 1 \,, \quad \sigma = 0 \,, \quad V_0 = 0 \,.
\end{equation}
Examining equations (\ref{eq62}) and (\ref{eq64}), we get
\begin{equation}
\label{eq91}
C_0 = - \frac{\partial V}{\partial q} \,, \quad C_2 = - \frac{\partial\,V/\partial\,t}{\partial\,V/\partial\,q} \,. 
\end{equation}
Also, instead of using expression (\ref{eq65}) for $C_2$, here it is more
useful to rewrite (\ref{eq63}) as
\begin{equation}
\label{eq92}
\frac{\partial\,C_{2}/\partial\,t}{\partial\,C_{2}/\partial\,q} = \frac{\partial\,V/\partial\,t}{\partial\,V/\partial\,q} \,. 
\end{equation}
The latter result simply states that $C_2$ and the potential are not
functionally independent, $C_2 = C_{2}(V)$. Thus, the potential satisfy
\begin{equation}
\label{eq93}
\frac{\partial V}{\partial t} + C_{2}(V)\frac{\partial V}{\partial q} = 0 \,.
\end{equation}
According to (\ref{eq67}), the compatible vector field can be expressed as  
\begin{equation}
\label{eq94}
{\bf v} = \frac{\partial V}{\partial q}\left(\frac{\partial}{\partial V} - \frac{1}{p - C_{2}(V)}\frac{\partial}{\partial p}\right) \,.
\end{equation}
By inspection, one of the characteristics of ${\bf v}$ is a function of
$p$ and $V$ only,
\begin{equation}
\label{eq500}
f = f(p,V) \,,
\end{equation}
where 
\begin{equation}
\label{eq501}
\frac{\partial f}{\partial V} - \frac{1}{p - C_{2}(V)}\frac{\partial f}{\partial p} = 0 \,.
\end{equation}
Using the second characteristic $g = t$, we have, by virtue of (\ref{eq91})
and (\ref{eq500}-\ref{eq501}),
\begin{equation}
\frac{df}{dg} = 0 \,,
\end{equation}
so that $f(p,V)$ is a first integral. 

For general $C_{2}(V)$, the solution of (\ref{eq501}) is not known, so
that Frobenius method only proves weak integrability. However, there is a
connection with the work of Giacomini \cite{Ref14}, in which it was made a
search of invariants depending on the momentum and the potential. Giacomini
has found exactly equations (\ref{eq93}) for the potential and (\ref{eq501})
for the invariant. Giacomini \cite{Ref14} as well as Bouquet and Lewis
\cite{Ref15} were able to construct explicit solutions for this system of
equations by choosing suitable functions $C_{2}(V)$. Nevertheless, we
have proven the interesting result that the Giacomini potentials are always
at least weakly integrable, for arbitrary $C_{2}(V)$, in the sense that they
admit a compatible vector field.
Also note that Giacomini's solution specified by $\rho = 1$, $\sigma = 0$
and $V_0 = 0$ does not excludes the case of invariants quadratic in
momentum, specified by $F = 0$. However, the intersection of the two
cases ($\rho = 1$, $\sigma = 0$, $V_0 = 0$ and $F = 0$) only yields the
trivial result that energy is conserved for a
time-independent potential. We do not pursue any further the use of
Frobenius method for Giacomini potentials, since it will not give new
results.

\subsection{An infinite class of weakly integrable potentials}

As a final example, we consider the choice
\begin{equation}
\label{eq95}
F(s) = s/k
\end{equation}
for the arbitrary function in equation (\ref{eq66}). Here, $s$ denotes an
arbitrary argument of $F$ and $k$ is a nonzero numerical constant. The
choice (\ref{eq95}) allows obtaining $Q$ globally using (\ref{eq66}), and
then $C_2$ by (\ref{eq65}). The result is
\begin{equation}
\label{eq96}
C_2 = \dot\sigma + \frac{(q - \sigma)}{\rho}\left(\dot\rho + (k + \frac{1}{\rho}\int^{t}\frac{dt'}{\rho^2})^{-1}\right) \,.
\end{equation}
It is possible to proceed to the solution of (\ref{eq64}) using this
function $C_2$, for arbitrary $\rho$, $\sigma$ and $V_0$. However, the
result is awkward and we content ourselves in taking
\begin{equation}
\label{eq97}
\sigma = 0 \,, \quad V_0 = 0 
\end{equation}
in the continuation. Thus, using (\ref{eq64}) and (\ref{eq96}), we find
the following equation for the potential,
\begin{equation}
\label{eq98}
\frac{\partial V}{\partial t} + \frac{q}{\rho}\left(\dot\rho + \frac{1}{\rho}(T + k)^{- 1}\right)\frac{\partial V}{\partial q} = - \frac{2\dot\rho}{\rho}\,V - \left(\frac{{\buildrel\cdots\over\rho}}{\rho} + \frac{3\dot\rho\ddot\rho}{\rho^2} + \frac{2\ddot\rho}{\rho^3}(T + k)^{- 1}\right)\frac{q^2}{2} \,,
\end{equation}
where
\begin{equation}
T = \int^{t}\frac{dt'}{\rho^2} \,.
\end{equation}
The solution for (\ref{eq98}) is 
\begin{equation}
\label{eq150}
V = \Gamma(t)\bar{q}^2 + \frac{1}{\rho^2}U(\bar{q}) \,,
\end{equation}
where
\begin{equation}
\Gamma(t) = -
\frac{1}{2\rho^2}\int^{t}\,\rho^{4}
\exp\left(2\int^{t'}\frac{dt''/\rho^2}{T + k}\right)\left(\frac{{\buildrel\cdots\over\rho}}{\rho} + \frac{3\dot\rho\ddot\rho}{\rho^2} + \frac{2\ddot\rho}{\rho^3}(T + k)^{- 1}\right)\,dt'
\end{equation}
and $U$ is an arbitrary function of 
\begin{equation}
\label{eq300}
\bar{q} = \frac{q}{\rho}\exp\left(- \int^{t}\frac{dt'/\rho^2}{T + k}\right) \,.
\end{equation}
The associated compatible vector field is 
\begin{equation}
\label{eq99}
{\bf v} = \frac{\partial}{\partial q} + \left(\frac{\dot\rho}{\rho} -
\frac{\rho\partial V/\partial q + \ddot\rho\,q}{\rho\,p -
\dot\rho\,q - (T + k)^{-1}q/\rho}\right)
\frac{\partial}{\partial p} \,,
\end{equation}
with $V$ given in (\ref{eq150}). 

To try to find the integral surfaces of ${\bf v}$, let 
\begin{equation}
\bar{p} = \rho\,p - \dot\rho\,q - (T + k)^{-1}q/\rho \,.
\end{equation}
In terms of $\bar{p}$ and $\bar{q}$ given in (\ref{eq300}), we express the
characteristic equation associated to ${\bf v}$ as
\begin{equation}
\label{eq101}
\bar{p}\frac{d\bar{p}}{d\bar{q}} + (T + k)^{-1}\exp\left(\int^{t}\frac{dt'/\rho^2}{T + k}\right)\bar{p} + \rho^{3}\ddot\rho\exp\left(2\int^{t}\frac{dt'/\rho^2}{T + k}\right)\bar{q} + 2\rho^{2}\Gamma\bar{q} + \frac{dU}{d\bar{q}} = 0 \,.
\end{equation}
Remembering that $t$ is simply a parameter in this equation
($t =$ cte. is one of the integral surfaces of the compatible
vector field), we can identify (\ref{eq101}) as an Abel's equation of
second type. Such Abel equation of second type is not solvable in terms of
elementary functions for
arbitrary $U(\bar{q})$. Hence, in general the class of potentials
(\ref{eq150})
is only weakly integrable. Finally, we observe that, if $\rho = 1$, one
obtains the Giacomini class of solutions. However, for $\rho \neq 1$ the
potential (\ref{eq150}) is new.

\section{Conclusion}

In this work, we have presented Frobenius method as an attractive strategy
for the integrability
analysis of one-dimensional time-dependent Hamiltonian systems, which may
be cast in the form
of three-dimensional dynamical systems. We have shown that the compatible
vector fields to be determined can be considered to depend only on a
function $C(q,p,t)$. The compatibility condition then leads to the basic
equation (\ref{eq16}), determining both $C$ and the potential. Starting
from some hypothesis on the functional dependence of $C$, we have been able
to derive some already known integrable potentials, namely the potential
with an invariant linear in momentum, Sarlet's  potential, the potential
for systems with a quadratic invariant and Giacomini's class of solutions.
However, the potential for the system with a quadratic invariant and
Giacomini potentials are only particular examples among a richer class,
defined by the solutions of (\ref{eq64}). These new solutions depends on
the arbitrary functions $\rho$, $\sigma$, $F$ and a further arbitrary
function arising after resolution of (\ref{eq64}), besides the trivial
function $V_{0}(t)$. In the final part of Section IV, we have constructed a
particular weakly integrable system contained in this family.

Frobenius method only assures weak integrability, without providing a first
integral in all cases. In this respect, there is a similarity with
Painlev\'e analysis \cite{Ref5}, which only points out when integrability
is likely to occur. Indeed, in Painlev\'e analysis the construction of an
invariant is a separate issue, to be addressed after the singularity
structure of the solutions is understood. Also note that, once a compatible
vector field is found, the Frobenius procedure for  construction of
invariants is entirely analogous to Lie's \cite{Ref3, Ref4} approach. The
difference between the two methods is that a compatible vector field is not
necessarily a generator of Lie symmetries.

Even if the existence of a compatible vector field does not allows the
construction of a first integral, if a first integral is known it is
given by Frobenius method. Indeed, let us make the transformation
\begin{equation}
\label{eq102}
C = - \frac{\partial J/\partial q}{\partial J/\partial p}
\end{equation}
on equation (\ref{eq16}), where $J = J(q,p,t)$ is an arbitrary function
of $p$, $q$ and $t$ such that $\partial J/\partial p \neq 0$. A little
calculation using the form (\ref{eq102}) shows that
\begin{equation}
\label{eq103}
{\bf u}(C) + C^2 + \frac{\partial^{2}V}{\partial q^2} = - \frac{1}{\partial J/\partial p}{\bf v}\left({\bf u}(J)\right) \,.
\end{equation}
Hence, if $J$ is an invariant, then the right hand side of (\ref{eq103})
is identically zero and the basic equation (\ref{eq16}) is satisfied by
the {\it ansatz} (\ref{eq102}). Moreover, the choice (\ref{eq102}) readily
implies that ${\bf v}(J) = 0$, showing that the first integral $J$ is
also a first integral of the compatible vector field. This shows that
{\it all} invariants for one-dimensional motion under a
time-dependent potential may be furnished by Frobenius method. In this
sense, Frobenius method has an universal character.

In most situations, no first integral is available {\it a priori}. In
these cases,
the crucial point for the effectiveness of the Frobenius approach is a
judicious choice of
the form of the compatible vector field. Clearly, it is possible that
other functional dependence of the coefficient $C$ on the compatible
vector field, different from the choices made in the present work, may
lead to useful results.
Another question that deserves attention is the extension of Frobenius
method and the notion of weak integrability to higher
dimensions. To conclude, we have seen that Frobenius method is a powerful
tool in integrability
analysis, and we expect that further results can be produced by its
systematic application.

\noindent
{\bf Acknowledgments}

\noindent
The author thanks the Laboratoire de Physique des Milieux Ionis\'es for the
hospitality while this work was carried out and the Brazilian agency
Conselho Nacional de Desenvolvimento Cient\'{\i}fico e Tecnol\'ogico for
financial support.

\end{document}